\documentclass[twocolumn,prl,showkeys,showpacs]{revtex4}
\usepackage{amsmath,amssymb}
\usepackage{graphicx,color}

\begin{document}

\title{Folding Kinetics of a Polymer}

\author{\v{S}t\v{e}p\'{a}n R\r{u}\v{z}i\v{c}ka}
\author{David Quigley}
\author{Michael P. Allen}
\affiliation{Department of Physics, University of Warwick, Coventry CV4 7AL, United Kingdom}

\date{\today}

\begin{abstract}
By simulating the first order globule-crystal transition of a flexible homopolymer chain, both by collision dynamics and Monte Carlo with non-kinetic moves, we show that the effective and the thermodynamic transition temperatures are different and we propose a way of quantifying the kinetic hindering. We then also observe that the top eigenvalue in the spectrum of the dynamical (contact or adjacency) matrix provides insight into the ensembles of folding and unfolding trajectories, and may be a suitable additional reaction coordinate for the folding transition of chain molecules.
\end{abstract}

\maketitle

Dynamical simulations of phase transitions in simple chain molecules are important for understanding the underlying microscopic folding mechanisms of polymers and proteins.  These simulations are usually computationally expensive, because the phases are often separated by a large free energy barrier, and the fluctuation leading to barrier crossing is a rare event on the molecular simulation timescale. Free energy barriers are usually studied by Monte Carlo (MC) simulation, and dynamical studies of barrier crossing are less common. Here, we use stochastic hard-sphere molecular dynamics, collision dynamics (CD) for short, to study the kinetics involved in the first-order crystallisation transition for a flexible homopolymer model, composed of bonded hard spheres with square-well nonbonded attractions, which has recently been intensively studied by MC \citep{Magee2006,Taylor2009,Taylor2009a,Seaton2010,Taylor2010}. By using CD, we avoid unphysical aspects of Monte Carlo (such as connectivity-altering moves improving sampling efficiency) and hence are able to observe the effects of kinetic bottlenecks in determining the rate, for a situation that mimics realistic dynamics.

In what follows, the polymer is represented by 128 hard spheres with two types of square-well potential for bonded and non-bonded pairs of monomers:
\begin{flalign}
u_{\text{b}}(r_{i, i+1}) & =
\begin{cases}
0 & \sigma < r_{i, i+1} < \chi_{\text{b}}\sigma \\
+\infty & \text{otherwise},
\end{cases} 
\label{eqn:ub}
\\
u_{\text{w}}(r_{ij}) & =
\begin{cases}	
+\infty & 0 < r_{ij} < \sigma \\
-\epsilon & \sigma < r_{ij} < \chi \sigma, \quad \quad \left|i-j\right| > 1, \\
0 & \chi \sigma < r_{ij}
\end{cases}
\label{eqn:uw}
\end{flalign}
where $r_{ij}$ is the distance between centers of two monomers $i$ and $j$, $\sigma$ is the diameter of the bead, $\epsilon$ is the depth of the square well, and $\chi$ and $\chi_{\text{b}}$ are the relative widths of the square well and the nearest-neighbour bond respectively. In the MC studies \cite{Taylor2009a} $\chi_{\text{b}}=1$, but for simple hard sphere dynamics, we choose a slightly larger value $\chi_{\text{b}} = 1.04$; we have verified that this makes very little difference to the equilibrium properties. \citet{Taylor2009,Taylor2009a} have determined that an all-in-one `protein-like' crystallisation from the expanded state occurs for $\chi\lesssim 1.06$, and a two-step `polymer-like' mechanism via a liquid-like globule, for $\chi\gtrsim 1.06$; we study the globule-crystal stage of the two-step process for $\chi$ values in this vicinity. All beads have equal mass $m$ which we take equal to unity. Throughout, we work in reduced units: $\sigma=1$, $\epsilon=1$ and $k_{\text{B}}=1$ (Boltzmann's constant) so $T=k_{\text{B}} T/\epsilon$.  Corresponding real values, for monomer beads corresponding to amino acids of the kind found in proteins, would be $m \approx 2 \times 10^{-25}$~kg, $\sigma \approx 6 \times 10^{-10}$~m, $\epsilon \approx 7 \times 10^{-22}$~J, and a time unit $\approx 10^{-11}$~s.

\begin{figure}[tp] 
   \centering 
   \includegraphics*[trim = 0 0 0 240]{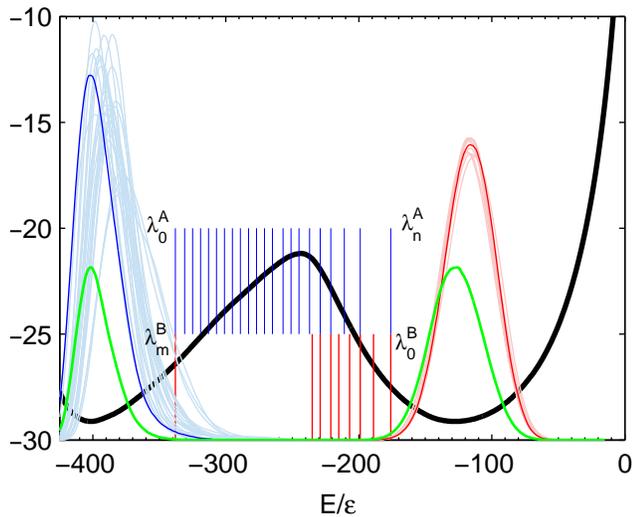} %
   \caption{\label{plot_double_hist} Free energy $F_C(E) = E - T \ln(g(E))$ (black) and canonical probability function $P_C(E) \propto g(E) \exp(-E/k_BT)$ (green) obtained from WL simulation. Vertical lines schematize the interfaces $\lambda_i^B$ in folding (red) and the interfaces $\lambda_i^A$ in unfolding (blue) directions. Displayed also are energy distributions of states in the crystalline phase (blue) and in the globule phase (red); these graphs must be scaled according to the folding and unfolding rates \citep{Valeriani2007} to be comparable with $P_C(E)$. The vertical axis of the probability function is not shown; these functions are normalized. The data are for the chain with $\chi = 1.07$, $T = 0.498$.} 
\end{figure}

We have simulated the above chain model with both the WL and CD methods. The WL Monte Carlo move set \cite{frenkel2002,landaubinderbook} consists of crankshaft, pivot, end-bridging, and regrowth moves; the latter two being connectivity altering. The regrowth move consists of regrowing up to 3 beads at either end of the chain, using a configuration-bias algorithm, and includes the possibility of reversing the chain. This MC move set was used with the WL algorithm as in \cite{Taylor2009} to iteratively approximate the density-of-states function $g(E)$, giving a well-sampled set of configurations across the whole energy range. In CD, free flight of the spheres occurs between elastic collisions in standard fashion \cite{allen1987,Rapaport2004}.  Collisions occur at each discontinuity in eqns~\eqref{eqn:ub}, \eqref{eqn:uw}. Additionally, thermal jolts reselect the velocities of individual atoms from the Boltzmann distribution and introduce a stochasticity into the dynamics. The time separation of the jolts has a Poisson distribution with mean time $\tau$ giving the strength of the coupling (typically $\tau = 0.1$ in reduced units). The WL simulations were used to determine the thermodynamic freezing temperatures, $T_f$, at which doubly-peaked canonical ensemble energy distributions $P_C(E)$ were observed, for a range of $\chi$ values. A typical example is shown in Figure~\ref{plot_double_hist}. Similar probability distributions were determined from the CD simulations. These confirmed a unique globule phase; we denote this as $B$ and identify a single energy distribution $P_B(E)$. However, CD simulations initialized in different realisations of the crystal phase did not explore as wide an energy range as the corresponding WL simulations. From this we infer that the crystal phase (at $T_f$) consists of a large number of basins with slightly different mean energies, separated by kinetic barriers which cannot be overcome (at these temperatures) on a simulation timescale without the use of unphysical MC moves. We therefore identify the \emph{crystal state} $A$ as being the state, that can be reached from other states within the distribution function $P_A(E)$ via constant temperature MC (including unphysical moves), which has the lowest mean energy. CD alone succeeded in sampling these low-lying basins only for $\chi \ge 1.05$.
\begin{figure}[tp] 
   \centering 
   \includegraphics*[trim = 0 0 0 240]{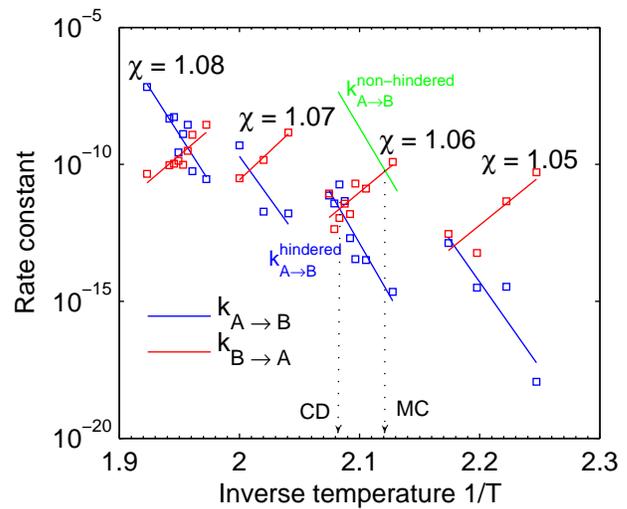}
   \caption{\label{ksvt1} Chevron plots with folding and unfolding rates computed by FFS. The intersections give the transition temperatures $T_f^{\text{CD}}$ estimated by collision dynamics. The schematic for $\chi = 1.06$ shows that kinetic hindering of unfolding in CD simulations can explain the observation that $T_f^{\text{CD}}>T_f^{\text{MC}}$. The schematic assumes that the hindering involved in the folding process is negligible.}
\end{figure}  

The order parameter $\lambda$ describing the qualitative difference between the globule and crystal state was chosen to be equal to the potential energy $E$ of the chain. To accelerate the sampling of fluctuations leading to the folding and unfolding transitions we used Forward Flux Sampling (FFS) \cite{Allenr2005, Allenr2009, escobedo2009}, which separates the phase space by $n$ hyperplanes orthogonal to the order parameter $\lambda$ and measures the probability flux through these planes. Here, the hyperplanes (also called interfaces) and the associated energies are denoted by the same symbol $\lambda_i$. FFS for our chain was performed in both directions, i.e.\ from globule to crystal, and from crystal to globule. This approach allows us to focus on kinetic effects associated with each direction separately, and is better suited to our dynamics than the backwards/forwards shooting approaches of alternatives such as Transition Interface Sampling, which are otherwise essentially equivalent \cite{Bolhuis2008}. FFS is also known to be relatively insensitive to the choice of order parameter. The rate from crystal to globule is given by:
\begin{equation}
k_{A \rightarrow B} = \frac{\left\langle \Phi_{A, \lambda^A_0} \right\rangle}{\left\langle h_A \right\rangle} \prod_{i=0}^{n-1} P(\lambda^A_{i+1} \mid \lambda^A_i) \:.
\label{eqn:kab}
\end{equation}
Here $\left\langle \Phi_{A,\lambda^A_0} \right\rangle$ is the probability flux through $\lambda^A_0$; state $A$ is defined by an interface $\lambda_A<\lambda^A_0$. $\left\langle h_A \right\rangle $ is the probability of being most recently in state $A$. The fraction $\left\langle \Phi_{A,\lambda^A_0} \right\rangle$ / $\left\langle h_A \right\rangle $ then simply represents the inverse of the average time needed to reach $\lambda^A_0$ from the first crossing of $\lambda_A$. $P(\lambda^A_{i+1} \mid \lambda^A_i)$ is the conditional probability of reaching the interface $\lambda^A_{i+1}$ from $\lambda^A_i$, which is given by the fraction of partial pathways started from $\lambda^A_i$ which reach $\lambda^A_{i+1}$ before they fall back to $\lambda_A$. The partial pathways are eventually connected into the full transition pathways starting in $A$ and ending in $B$. The rate $k_{B \rightarrow A}$ from globule to crystal is given analogously. In this way, we gathered 8192 folding and unfolding transition pathways by CD simulation for each $\chi$ and $T$, choosing temperatures in the vicinity of the transition temperatures determined by WL.

The interfaces in FFS are positioned as follows. Energies $\lambda_0^A$ and $\lambda_0^B$ are chosen such that the ranges $\lambda<\lambda_0^A$ and $\lambda>\lambda_0^B$ capture 99.9\% of the corresponding integrated densities $P_A$ and $P_B$.  The boundaries of $A$ and $B$ are then defined as $\lambda_A \equiv \lambda^A_0-50$ and $\lambda_B \equiv \lambda^B_0+50$. To reconstruct the full pathways we define an extra plane as $\lambda^A_n \equiv \lambda_0^B$. Interface $\lambda_{n-1}^A$ is placed close to the isocommittor, specifically such that $P(\lambda^B \mid \lambda^A_{n-1}) = 0.9$. The remaining interfaces  $\lambda^A_{1},...,\lambda^A_{n-2}$ are chosen according to the optimization scheme of Ref.~\cite{escobedo2009}. An analogous procedure applies to interfaces $\lambda^B_{1},...,\lambda^B_{m}$.

To avoid undersampling of $\lambda_0^B$, 32 random disordered chains are equilibrated, and $\lambda_0^B$ is sampled in parallel starting from these different chains. Similarly, CD simulations started from different points in $\lambda_m^B$ gave $P_A^{(i)}(E)$, $i=1,...,256$. $\lambda_0^A$ is then sampled in parallel starting from a configuration belonging to the distribution $P_A(E)$ having the lowest mean energy, and from 31 other configurations which are separated from this state by a large number of MC moves, including connectivity-altering moves, conducted at constant temperature.
\begin{figure}[tp] 
   \centering 
   \includegraphics*[trim = 0 0 0 240]{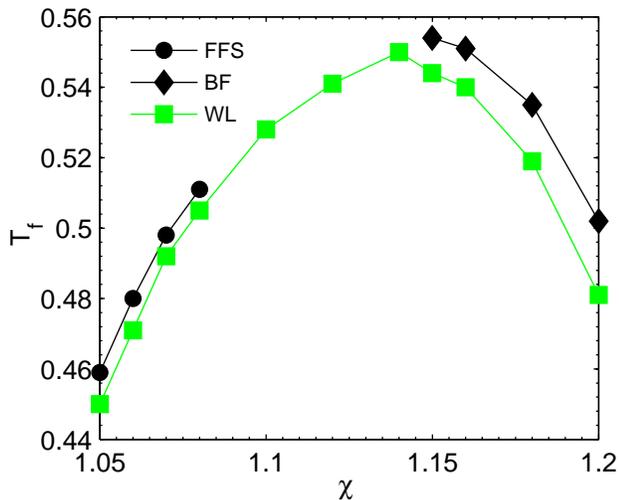} %
   \caption{\label{pdiam} Phase diagram with freezing temperatures determined from chevron plots ($T_f^{\text{CD}}$ , FFS and BF) and by Maxwell construction ($T_f^{\text{MC}}$, WL). The errors are smaller than the symbol sizes.}
\end{figure}
 
Transition temperatures $T_f^{\text{CD}}$, defined dynamically by the equation $k_{A \rightarrow B}=k_{B \rightarrow A}$, are obtained from the FFS simulations via a so-called chevron plot, Fig.~\ref{ksvt1} (based on the assumption of Arrhenius-like behaviour). These were slightly, but systematically, higher than the WL transition temperatures $T_f^{\text{MC}}$ by an amount $\Delta T = T_f^{\text{CD}}-T_f^{\text{MC}}= 0.005$--$0.012$  (Fig.~\ref{pdiam}). For the larger values of $\chi$, barrier crossing could be observed directly, and rate constants were calculated by brute force (BF) simulation. Transition temperatures obtained in this way were again comparable to the temperatures obtained in MC simulation, but were also systematically higher. This suggests that the discrepancy is due to a real dynamical (kinetic) effect rather than any deficiency of the FFS algorithm itself. The most likely explanation is that CD, hindered by metastable basins acting as kinetic traps, yields significantly lower unfolding rates than the MC simulation, which escapes the traps with the aid of non-kinetic connectivity-altering moves.  Fig.~\ref{ksvt1} indicates schematically the extent to which the unfolding regression line must be shifted to give the observed shift $\Delta T$; this could provide a quantitative measure of the kinetic hindering. 

The remaining part of the paper points out a remarkable property the \emph{dynamical matrix} of our polymer system in the vicinity of the free energy maximum. We focus on the chain with $\chi = 1.07$ and $T = 0.498$, but similar results were obtained for other parameters, becoming even more distinct with decreasing $\chi$. The dynamical matrix of our chain is defined as follows:
\begin{flalign}
\Gamma_{ij} & =
\begin{cases}
 -1 & \text{ if } i \neq j \text{ and } r_{ij} \le \chi\sigma, \\
0 & \text{ if } i \neq j \text{ and } r_{ij} > \chi\sigma, \\
-\sum_{i,i \neq j} \Gamma_{ij} & \text{ if } i = j.
\end{cases}
\label{eqn:dyn_mat}
\end{flalign}
It is closely related to the contact (or adjacency) matrix, whose elements are unity for atom pairs within interaction range, and zero otherwise. Contact matrices have been used to describe the equilibrium structure of proteins in terms of amino acid contacts \citep{Vendruscolo1999,England2003}; as described by \citet{Bahar1998} and \citet{ Sadoc2005} this idea may be extended, through the dynamical matrix, to give a simple network model of vibrational motions. Here we suggest that the topology of the interactions in the chain around the critical point of collapse, described by $\Gamma$, may be linked to the dynamical effects that we have observed.

\begin{figure}[tp] 
   \centering 
   \includegraphics[width=6cm, viewport = 30 0 220 220]{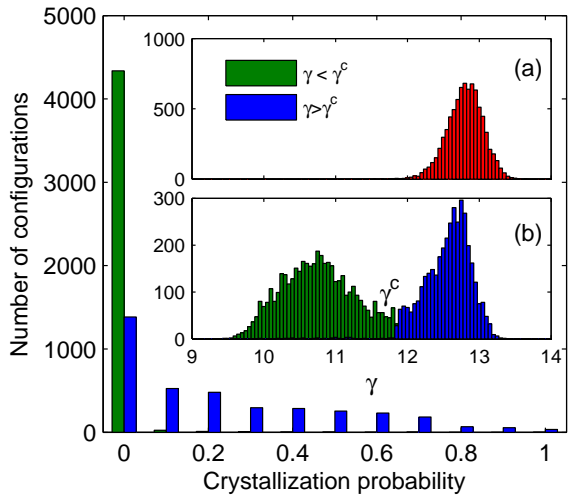} 
   \caption{\label{egn_pbr} Configurations at $\lambda_{n-1}^B$ with largest eigenvalue $\gamma$ of the dynamical matrix lower than the critical value ($\gamma^c = 11.9$) have almost no chance to crystallize. Insets: (a) Unimodal distribution $a(\lambda^A;\gamma)$ at the surface $\lambda^A = -235$ sampled by pathways started in $A$. (b) Bimodal distribution of $b(\lambda^B;\gamma)$ at the same surface ($\lambda_{n-1}^B = -235$) but sampled by pathways started in $B$.}
\end{figure}

The largest eigenvalue of $\Gamma$ will be denoted as $\gamma$. Let $a(\lambda;\gamma)$ be the probability distribution of $\gamma$ at an interface $\lambda$ sampled by pathways started from $A$, and  $b(\lambda;\gamma)$ the same quantity but sampled by pathways started from $B$. We found that $a(\lambda;\gamma)$ is unimodal (approximately Gaussian) at all interfaces, with the mean value growing with $\lambda$, and that the conformations at $\lambda$ with large $\gamma$ are more likely to crystallise. The distribution $b(\lambda;\gamma)$ at interfaces far enough from the isocommittor is also unimodal with similar properties, but becomes bimodal at the interfaces $\lambda$ close to the isocommittor. The critical value separating these two modes is denoted as $\gamma^c$. The insets of Fig.~\ref{egn_pbr} show that pathways started from $A$, and reaching these energies, do not sample the population of low-eigenvalue states. The microscopic reversibility of our dynamics then implies that folding transition pathways must cross $\lambda_{n-1}^B$ at $\gamma > \gamma^c$. Indeed, the probability analysis in Fig.~\ref{egn_pbr} shows that pathways started at $\lambda_{n-1}^B$ with $\gamma < \gamma^c$ have almost no chance to reach $A$. The difference between the distributions $a(\lambda;\gamma)$ and $b(\lambda;\gamma)$ confirms our suspicion that the forward-going and reverse-going ensembles of reactive trajectories are not identical (due to kinetic hindering), and that the top eigenvalue of the dynamical or contact matrix may be able to discriminate between these families of trajectories. Incidentally, the \emph{equilibrium} distribution of $\gamma$ at this energy, obtained by WL, is very similar to the nonreactive $\gamma<\gamma^c$ portion of Fig.~\ref{egn_pbr}(b). Similar observations apply to other eigenvalues near the top of the spectrum of $\Gamma$. It is significant that typical $\gamma>\gamma^c$ configurations appear to have a compact crystal nucleus with attached chains or loops, while for $\gamma<\gamma^c$ the same number of interactions are typically arranged in a single, less well ordered, cluster.

It is worth mentioning that two structures on surfaces $\lambda$ close to the isocommittor have also been identified by \citet{Taylor2010} using the radius of gyration ($R_g$) as a second reaction coordinate. This result was also confirmed here. The correlation between $R_g$ and $\gamma$ was tested and found to be only weak. An analysis similar to that in Fig.~\ref{egn_pbr} showed that $\gamma$ has significantly better predictive properties than $R_g$.

Why do we believe that the dynamical matrix, and the associated contact matrix, deserve further study? As mentioned above, they give a rather general connection between the topology defined by the interactions within a chain configuration and its dynamical evolution, in the approximation of an elastic network model. This has not only been used in the discussion of proteins to identify vibrational modes of oscillations \citep{Bahar1998, Sadoc2005}, but also in the definition of nodes and an order parameter (a distance between nodes) in dynamical network models of the folding process itself  \citep{Lois2010}. The dynamical matrix is also being used for the description of glassy structures in colloidal systems \citep{Chen2011}.  Most recently, the contact matrix of an atomic cluster has been used as a generator of order parameters for metadynamics simulations \citep{Pietrucci2011}. Our results clearly reinforce the view that the dynamical matrix is a simple object capturing successfully important topological or vibrational features of various interacting systems.

To summarize, the transition of the homopolymer chain from the disordered globule to the crystal state has been simulated by dynamical forward flux sampling and brute force simulation. The results gave strong evidence that kinetic effects play an important role in the determination of the effective transition temperature. We then showed that the eigenvalues of the dynamical matrix yield further important information regarding the forward and reverse trajectories in the folding transition which complement the potential energy as an order parameter. 

\begin{acknowledgments}
Computer facilities were provided by the Centre for Scientific Computing at the University of Warwick. We thank Dr. Rosalind Allen, Prof. Kurt Binder, Dr. Ellak Somfai, and Adam Swetnam for useful discussions.
\end{acknowledgments}

\bibliography{biblioa}

\end{document}